\documentclass[journal=enfuem,manuscript=article]{achemso}

 \usepackage{float}
 \usepackage[labelfont=bf,labelsep=period]{caption}
 \usepackage{sectsty}
 \usepackage[utf8]{inputenc}
 \usepackage{array}
 \usepackage{geometry}
 \usepackage{setspace}
 \usepackage{xkeyval}
 \usepackage{gensymb}
 \usepackage{xcolor}
 \usepackage{natbib}
 \usepackage{comment}
 \usepackage{soul}
 \usepackage{nomencl}
 \usepackage[shortlabels]{enumitem}
\makenomenclature

\renewcommand{\nompreamble}{The list describes several symbols used within the body of the document}
 
\author{Gilmar F. Arends}
\author{John M. Shaw}
\author{Xuehua Zhang}
\affiliation{Department of Chemical and Materials Engineering, University of Alberta, Alberta T6G 1H9, Canada}
\email{xuehua.zhang@ualberta.ca}

\title{How fast do microdroplets generated during liquid-liquid phase separation move in a confined 2D space?}
\DeclareUnicodeCharacter{2212}{-}

\begin{document}

\nomenclature{$Ma$}{Marangoni number}
\nomenclature{$d$}{Position along data acquisition line perpendicular to triangular protrusion}
\nomenclature{$d_{final}$}{Final position of water-rich droplets where they merge at boundary between zone 2 and 3}
\nomenclature{$x$}{Position along x-direction}
\nomenclature{$x_0$}{Initial position at triangular protrusion tip}
\nomenclature{$w_{eth23,3}$}{Ethanol mass fraction at boundary between 2 and 3 from zone 3 side}
\nomenclature{$w_{eth23,2}$}{Ethanol mass fraction at boundary between 2 and 3 from zone 2 side}
\nomenclature{$w_{eth12,2}$}{Ethanol mass fraction at boundary between 1 and 2 from zone 2 side}
\nomenclature{$l$}{Droplet displacement}
\nomenclature{$I(d)$}{Mean gray intensity as function of position $d$ along triangular protrusion}
\nomenclature{$I_0$}{Mean gray intensity at side channel position $x=0$}
\nomenclature{$I(x)$}{Mean gray intensity as function of position $x$}
\nomenclature{$I_{di}$}{Mean gray intensity at position $d=0$, starting along fixed y-position at the center of a triangular zone}
\nomenclature{$I_{norm}(d)$}{Normalized intensity across triangular protrusion as function of position $d$}
\nomenclature{$I_{df}$}{Mean gray intensity at final position $df$, at the water-rich side of a triangular zone}
\nomenclature{$t$}{Time}
\nomenclature{$a(x)$}{Fitting parameter for normalized intensity profile at fixed x-position and time t}
\nomenclature{$w_{eth}$}{Ethanol mass fraction}
\nomenclature{$w_{eth12,1}$}{Ethanol mass fraction at the boundary 1 between zone 1 and 2}
\nomenclature{$w_{eth12,2}$}{Ethanol mass fraction at the boundary 2 between zone 1 and 2}
\nomenclature{$w_{eth23,2}$}{Ethanol mass fraction at the boundary 2 between zone 2 and 3}
\nomenclature{$w_{eth23,3}$}{Ethanol mass fraction at the boundary 3 between zone 2 and 3}
\nomenclature{$\Delta \sigma$}{Difference in surface tension}
\nomenclature{$R$}{Droplet radius}
\nomenclature{$\mu$}{Viscosity of liquid surrounding the droplet surface}
\nomenclature{$D$}{Diffusion coefficient of ethanol within liquid surrounding the droplet surface}
\nomenclature{$\Delta \rho_{eth}$}{Ethanol mass concentration change in surrounding liquid}
\nomenclature{$\rho$}{Surrounding liquid density}
\nomenclature{$D$}{Droplet displacement scaling with time}
\nomenclature{$\omega$}{Triangular protrusion angle}
\nomenclature{$\theta$}{Droplet propulsion angle}

\begin{abstract}

How liquid transport occurs in confined spaces is relevant to many industrial and lab-scale processes, ranging from enhanced oil recovery to drug delivery systems. In this work, we investigate propelling microdroplets that form from liquid-liquid phase separation in a quasi-2D chamber, focusing on the direction and speed of microdroplets in response to local composition gradients. The confined ternary solution in our experiments comprises a model oil (the main one being octanol), a good solvent (ethanol) and a poor solvent (water). It is displaced by water. Depending on the initial solution composition, water-rich or oil-rich microdroplets, $\sim$ $1/4-1/3$ of the height of a narrow and wide microchamber, form and move spontaneously as the ternary solution mixes with and is displaced by water diffusing from a deep side channel. Microdroplet movement is followed in situ using high-speed bright-field imaging or fluorescence imaging when the solution is doped with a dye. Local ethanol composition gradients are estimated from the variation of fluorescence intensity in the local continuous liquid surrounding of the mobile microdroplets. From phase separation of the ternary solution with high oil concentration, mobile oil-rich microdroplets form in a water-rich zone, accompanying the formation of water-rich microdroplets in an oil-rich zone. The fast movement of oil-rich microdroplets induces directional flow transport that mobilizes water-rich microdroplets close to the water-rich zone. Regardless of the initial composition of the solution, displacement of oil-rich microdroplets extends linearly with time. The average microdroplet speed increases with the initial oil concentration in the ternary solution. The fastest speed of oil-rich microdroplets observed in our experiments is $\sim$150 $\mu m/s$ along the surface of a hydrophobic wall. The presence of a sharp ethanol composition gradient is thought to be the primary driving force for the fast movement of oil-rich microdroplets in confinement. Our results demonstrate the potential of enhancing liquid transport in confinement through composition gradients arising from phase separation.

\end{abstract}

\section{Introduction}

The creation and movement of microdroplets in confinement hold promise for improved fluid transport in enhanced oil recovery processes\cite{Perazzo2018, Zhou2019, Tagavifar2017}, novel chemical analysis techniques \cite{Hubner2020}, drug and gene delivery systems \cite{Tomeh2020}, smart colloid transportation \cite{Wang2019, Zwanikken2016, Moller2017, Das2017, Brady2011, Selva2012}, and chemical synthesis \cite{Grommet2020, Fallah2014}. Recently, the autonomous motion of microdroplets has attracted research interest, due to the potential to imitate numerous forms of life-like and robotic behaviours \cite{Cejkova2014, Liu2019, Wang2019}. For example, droplets may display individual \cite{Lagzi2010, Jin2017} or collective chemotactic behaviour \cite{Liebchen2018,Thutupalli2011}, controlled clustering \cite{Krueger2016} and the capability of carrying payloads through complex geometries \cite{Jin2017, Izri2014, Cejkova2014}. However, the motion of droplets arising from confined liquid-liquid phase separation multicomponent mixtures is an example requiring greater understanding.
 
The common mechanisms for the autonomous motion of droplets involve a Marangoni effect induced by surface tension gradients \cite{Yang2018, LiY2019, Liu2019, Cira2015, Frenkel2017, Das2017, Shin2016, Kar2015}. Droplets can travel over a long distance as long as the driving force from the unequilibrated chemical constituent can be sustained. A dimensionless Marangoni number ($Ma$) provides a quantitative comparison between the rate of transport by Marangoni flow versus the rate of transport by diffusion \cite{Lohse2020}. A large $Ma$ means transport is dominated by a surface tension gradient. A low $Ma$ means transport is dominated by diffusion instead.

Microdroplet motion can potentially reach high speeds and in designated directions by manipulating surface tension gradients around them. These gradients may be generated by many chemical approaches, including changes in the local pH \cite{Moller2017, Ban2013}, surfactant concentration \cite{Jin2017}, the concentration of chemically reactive species or products \cite{Tanaka2021}, or combinations of the above. For example, in the work by Thutupalli et al. \cite{Thutupalli2011}, a reactant-containing water droplet is in contact with an oil phase containing a surfactant. The reactant in the droplet reacts with the surfactant at the droplet surface, leading to increased local surface tension. A Marangoni flow along the droplet surface propels the droplet and replenishes the reacting droplet surface with fresh surfactant. The self-sustaining droplet motion continues until the reactant in the droplet is exhausted. 

Concentration gradients can also arise through spontaneous phase separation in multicomponent mixtures. A simple yet representative process is the formation of droplets via the Ouzo effect, where micro-sized droplets spontaneously form when a ternary mixture (e.g. ethanol, water and oil) is diluted with a poor solvent (water) \cite{Vitale2003, Li2020, Zemb2016, Solans2016}. Concentration gradients in ternary ouzo solutions can be sufficient to drive a macroscopic drop to oscillate or jump \cite{lu2017nanoscale,LiY2019}. Li et al. reported the fascinating jump and sudden death of a sub-millimetre oil drop in vertically stratified ethanol + water mixture \cite{LiY2019}. The sharp ethanol concentration gradient induces a Marangoni flow that propels the droplet to repeatedly jump up to about 10 times the droplet diameter, which then falls under gravity. The cycle of jumping and falling repeats at a frequency of 2-3 Hz and lasts for approximately 30 minutes until the concentration gradient decays \cite{LiY2019}. Sudden death occurs when the Marangoni flow is not able to sustain the jumping behaviour and buoyancy dominates.

The speed of microdroplets, in confinement, reported in the literature ranges from 10 to $10^{4}$ ${\mu}m/s$ \cite{Herminghaus2014, Izri2014, Ban2013}, depending on droplet size and the driving force of the autonomous motion. What controls the speed of propelling droplets in confined spaces is an important and unresolved question. In confined spaces, concentration gradients, once created, lasts for a long time because mixing and transport occur through diffusion. Our latest work showed that large concentration gradients from the Ouzo effect could propel microdroplets \cite{Lu2017}, and lead to a fast replenishing flow that sustains further droplet formation and propulsion from phase separation \cite{Zhang2020}. Phase separation may also produce spatially segregated oil-rich and water-rich zones in confinement. The spatial arrangement, relative amount and composition of subphases formed are closely linked to the compositions of the mixture undergoing phase separation and the penetrating liquid through the underlying phase diagram \cite{Arends2021}. However, it remains unclear how the initial composition of the ternary mixture and local concentration gradients affect the dynamics of droplets propelled by the ouzo effect. 

In this work, we investigate the effect of initial solution composition and local composition gradients on the dynamic behaviour of microdroplets by diluting ternary mixtures with water in a quasi-2D confined space. The ternary mixture consists of a model oil (1-octanol butylparaben, HDODA or Oleic acid), ethanol and water. The chemical concentration gradients are estimated from the local fluorescence intensity. The maximum speed of the oil-rich microdroplets was found to reach $\sim$150 ${\mu}m/s$ due to a sharp gradient in the ethanol concentration. Our findings provide a better understanding of the concentration gradient-driven mobility of droplets in a confined space, with one dimension being only 3-4 times the diameter of the microdroplets. Developing methods to improve mass transport in confinement, such as by manipulating composition gradients to induce the autonomous motion of droplets, would be beneficial to a range of technological applications, such as liquid extraction from porous media, pharmaceutical delivery systems, chemical diagnostic devices, or separation by porous membranes.

\newpage
\section{Experimental Section}

\subsection{Chemicals and solution preparation}

\begin{figure}[H]
	\includegraphics[trim={0cm 0cm 0cm 0cm}, clip, width=0.9\columnwidth]{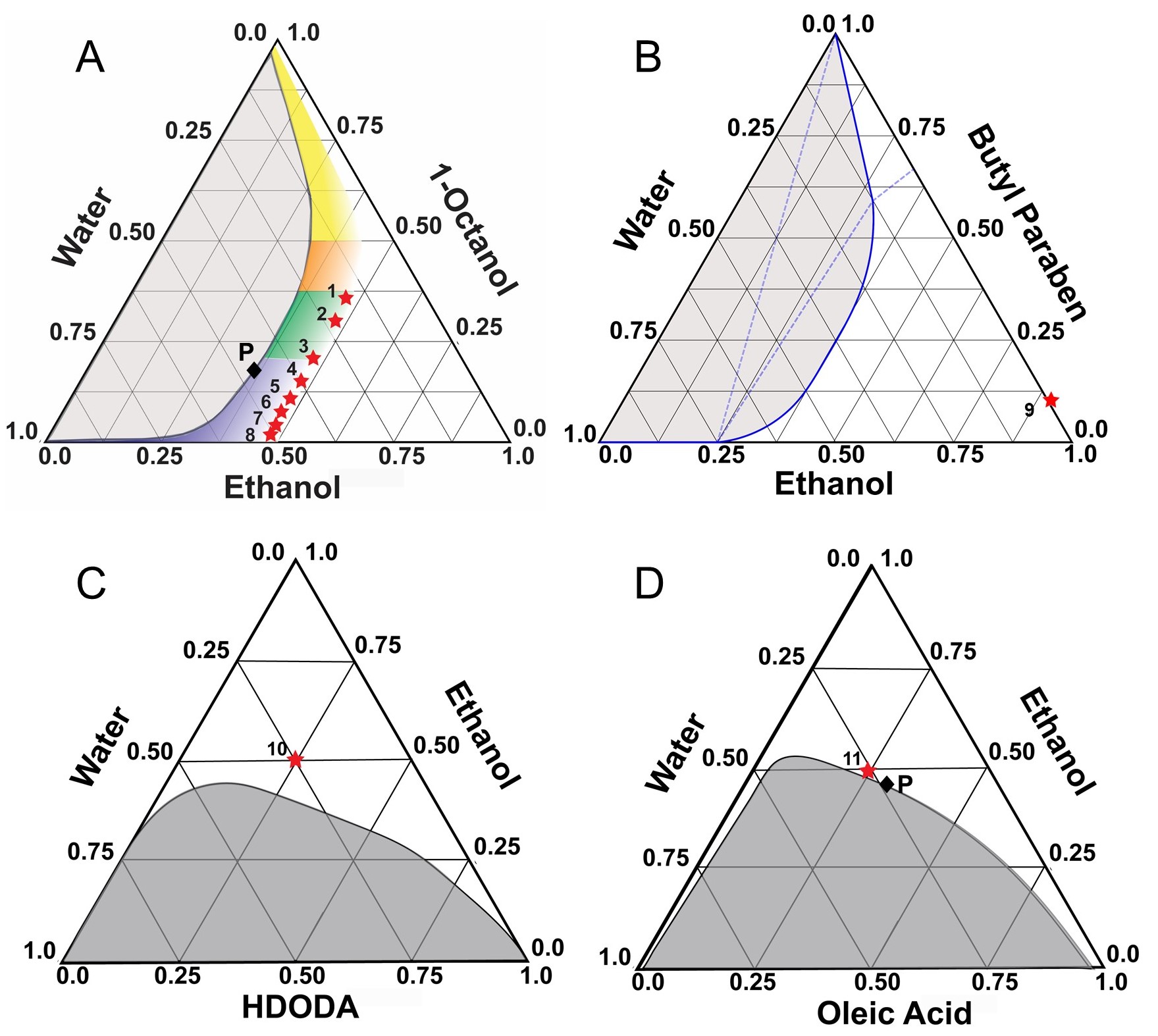}
	\caption{(A) Ternary phase diagram with composition regimes for the 1-octanol, ethanol and water ternary in terms of mass fraction. The initial solution compositions are indicated with a star and annotated with the experiment number same as in Table \ref{Experimental Compositions Table}. Reproduced from \textit{Energy Fuels} \textbf{2021}, 35, 6, 5194–5205 (\cite{Arends2021}). Copyright 2021 American Chemical Society. P indicates the Plait point ($L_1=L_2$). Additional phase diagrams are for: (B) butylparaben, ethanol, water in mass fractions \cite{yang2015phase}; (C) HDODA, ethanol and water in mass fractions \cite{li2019controlled}; (D) Oleic acid, ethanol and water in mass fractions \cite{rahman2003ternary}.
	}
	\label{ternarydiagrams}
\end{figure}

Solution A, the solution present in the 2D microchannel, comprised a mixture of either 1-octanol (ACS grade $>$95\%, Fischer Scientific) or butylparaben ($>$99\%, Sigma-Aldrich), ethanol (Histological grade, Fischer Scientific) and water (Milli-Q). Additional experiments were performed with 1,6-hexanediol diacrylate HDODA ($>$98.5\% total reactive esters, Fisher Scientific - Alfa Aesar) and oleic acid ($>$99.0\%, Fisher Scientific - Alfa Aesar) as model oils in solution A. The detailed compositions of solution A, in the experimental matrix, are listed in Table \ref{Experimental Compositions Table} and are labelled in the ternary phase diagrams in Figure \ref{ternarydiagrams}. 

A lipophilic stain Nile Red (Fischer Scientific) was added to solution A, at $1.0 \times 10^{-5} M$, so that the oil-rich phase could be identified from the fluorescence intensity of the dye after liquid-liquid phase separation. This dye partitions 99.9\% to the octanol-rich subphase, and the fluorescence intensity from water is zero \cite{Wang2012}.

\begin{table}[H]
\captionsetup{font = {small}}
\caption{Compositions of solution A in the experiments. Phase diagrams annotated with the compositions comprise Figure \ref{ternarydiagrams}.}
\centering
\begin{tabular}{|c| c| c| c|}
\hline
Composition & Model Oil & Ethanol & Water \\ &Mass \% &Mass \% &Mass \%\\
\hline
1 & 35 (1-Octanol) & 48 & 17 \\
\hline
2 & 30 & 48 & 22 \\
\hline
3 & 20 & 48 & 32 \\
\hline
4 & 15 & 48 & 37 \\
\hline
5 & 10 & 48 & 42 \\
\hline
6 & 8 & 47 & 45 \\
\hline
7 & 5 & 48 & 47 \\
\hline
8 & 2 & 48 & 50 \\
\hline
9 & 10 (Butylparaben) & 90 & 0 \\
\hline
10 & 25 (HDODA) & 50 & 25 \\
\hline
11 & 24 (Oleic acid) & 49 & 27 \\
\hline
\end{tabular}
\label{Experimental Compositions Table}
\end{table}

\subsection{Setup and wall surface properties of the quasi-2D microchamber}

\begin{figure}[H]
\centering
	\includegraphics[trim={0cm 0cm 0cm 0cm}, clip, width=0.7\columnwidth]{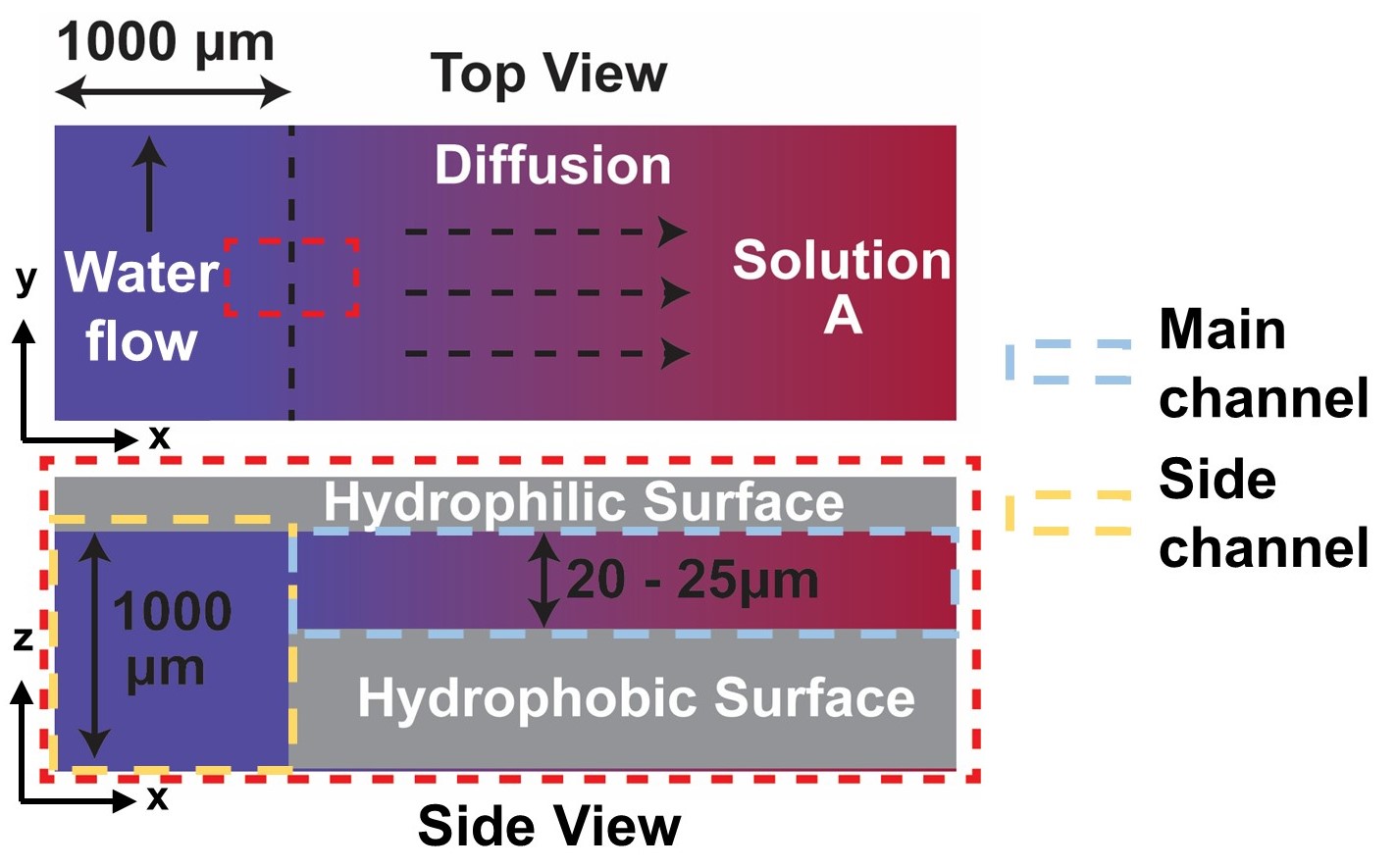}
	\caption{Top view and cross-section sketches of the main and side channels of the microchannel. The ternary solution A fills the narrow and wide main channel. The poor solvent water flows in the deep side channel and diffuses transversely into the main channel. Adapted from \textit{Energy Fuels} \textbf{2021}, 35, 6, 5194–5205 (\cite{Arends2021}). Copyright 2021 American Chemical Society.
	}
	\label{experimental setup}
\end{figure}

A microchannel was used to provide a quasi-2D space for diffusive mixing between solutions A and B. The entire chamber consisted of a polycarbonate base (8.5 $cm$ $\times$ 4.5 $cm$ $\times$ 1 $cm$), a hydrophilic glass slide (Fisherbrand Microscope Slide) top plate, with a 1 $mm$ silicon spacer in between. A silicon wafer was placed on top of the polycarbonate base and was functionalized to render it hydrophobic. The main microchannel height was 20 to 25 ${\mu}m$ (Figure \ref{experimental setup}). The side channel height was 1 $mm$. The glass slide top plate was clamped to the base with binder clips. Inlet and outlet tubing were connected to the polycarbonate base.

The silicon wafers were washed with piranha solution consisting of 70\%-volume $H_2SO_4$ (ACS Plus grade, Fischer Scientific) and 30\%-volume $H_2O_2$ (30\% ACS grade, Fischer Scientific) at 85$\degree$ C for 20 minutes. After the initial cleaning, the silicon wafers were sonicated in a bath of water and then in ethanol sequentially for 15 minutes. The wafers were hydrophobized with octadecyltrichlorosilane (OTS) following an established procedure \cite{ZhangX2008}. In brief, 0.5 vol.\% OTS in hexane solution was used to soak wafers for 12 hours at ambient conditions. Wafers were then sonicated in hexane, acetone, and ethanol for 10 minutes each.  After the hydrophobization, the wafers were cut to a suitable dimension before use. Before experiments, all parts of the flow channel were sonicated in water and ethanol for 10 minutes each.

\subsection{Visualization of solvent mixing in the microchannel}

The events occurring while solution B was pumped (at 12 mL/hr) into the chamber using a syringe pump (Fisherbrand Single Syringe Pump) were captured with a video camera (Nikon DS-Fi3) connected to an optical microscope (Nikon Eclipse Ni-U). Visible light images were captured using an auto-exposure setting at 10 to 15 frames per second (FPS) with a fixed resolution of 2880 $\times$ 2048 pixels. Fluorescence videos were captured at 1 FPS due to longer exposure times needed. A high-speed camera (Photron Fastcam Mini UX100) was used to capture short time-scale events. These were captured at 500 or 1000 FPS over 4.5 seconds at a resolution of 1280 $\times$ 1024 pixels. 

\subsection{Droplet tracking and data analysis}

The videos were processed and converted to 8-bit grayscale image sequences using ImageJ (Fiji package) software. The contrast and brightness levels were sometimes adjusted. Droplet tracking was performed manually, using high-speed image sequences with an ImageJ plugin, at intervals of 0.02 s. The spatial resolution was 0.24 $\mu$m/px for the Nikon DS-Fi3 camera and 1 $\mu$m/px for the Photron Fastcam Mini UX100 camera at a microscope magnification of 10$\times$. The tracking data from ImageJ was saved as a CSV file, which included frame and position data in pixels. 

At different solution compositions, there were distinct features to be analyzed. Oil-rich and water-rich droplets were tracked as they nucleated or became visible in the field of view. Tracking was terminated when droplets came to a stop or when they merged with nearby phase domains. Droplet displacement ($l$) was calculated from the x and y positions as a function of time ($t$) using the first entries in the data set ($t=0$) as the initial positions where:

\begin{equation}
l(t)=\sqrt{(\Delta x^2 + \Delta y^2)}=\sqrt{[(x(t)-x(t=0)]^2 + [y(t)-y(t=0)]^2}
\label{dropletdisplacement}
\end{equation}

\subsection{Image analysis for local concentration gradients}

The Marangoni number compares the transport related to the stress imposed by concentration gradients versus the transport related to diffusion. The dimensionless Marangoni number was used to relate droplet mobility to ethanol composition gradients within a given phase. The dimensionless Marangoni number $Ma$ for a droplet within 2D space is expressed as \cite{Lohse2020}: 

\begin{equation}
Ma=\Delta \sigma \left ( \frac{R}{\rho \nu D} \right ) =\Delta \sigma \left ( \frac{R}{\mu D} \right )  
\label{marangoninumber}
\end{equation}
\noindent
where $R$ is the droplet radius or characteristic length in $m$, $\mu$ is the viscosity of the surrounding liquid in $Pa\cdot s$, and $D$ is the diffusion constant in $m^{2}/s$. Due to the small length scale of the droplets, the viscosity is taken as a constant. The surface tension gradient is expressed as a function of x-position \cite{Lohse2020}:

\begin{equation}
\Delta \sigma \approx \frac{\partial \sigma}{\partial {\rho}_{eth}} \Delta {\rho}_{eth} \approx \frac{\partial \sigma}{\partial w_{eth}} \Delta w_{eth} (x)
\end{equation}

\noindent
where $\sigma$ is the surface tension as a function of ethanol composition ($w_{eth}$). The surface tension values were approximated, based on local compositions, using the UNIFAC thermodynamic model within the software package Symmetry.

\begin{figure}[H]
	\includegraphics[trim={0cm 0cm 0cm 0cm}, clip, width=0.80\columnwidth]{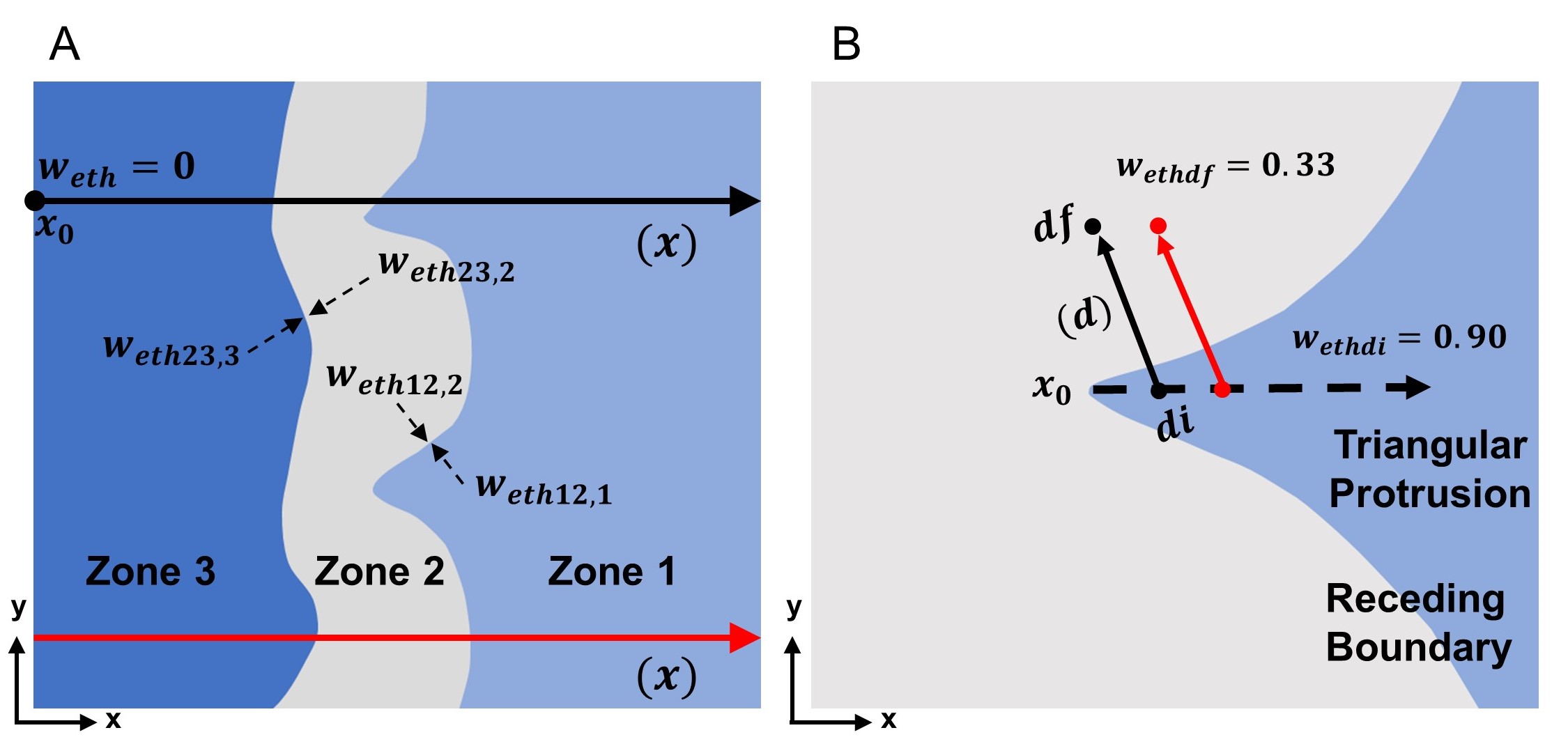}
	\caption{Fluorescence intensity characterization for two configurations of the liquid mixtures. (A) Fluorescence intensity values were taken as a function of x-direction. The two horizontal lines illustrate intensity data as a function of x-position at two y-values. At $x_0$ the ethanol fraction ($w_{eth}$) is taken to be 0. Ethanol fraction at each boundary location is labelled with zone numbers. $w_{eth12,1}$ represents ethanol mass fraction at boundary between zone 1 and 2 from zone 1 side. $w_{eth12,1}$ has solution A ethanol composition, $w_{eth12,2}$ water-rich phase composition, $w_{eth23,3}$ oil-rich phase composition. (B) Fluorescence intensity values taken as function of position ($d$) on parallel lines along the x-direction within and adjacent a triangular feature on a boundary. $di$ and $df$ are the initial and final positions of intensity measurements with ethanol fraction $w_{ethdi}$ and $w_{wthdf}$. Fluorescence intensity measurements at a triangular feature start from the tip location at $x_0$ and continue towards the positive x-direction.
	}
	\label{acquisition}
\end{figure}

At a given time in the main channel, solution A, water, and subphases from phase separation evolve into two different configurations: a three-zone configuration and a triangular protrusion of solution A behind the receding boundary as sketched in Figure \ref{acquisition}. The local mass fraction of the good solvent (ethanol) was estimated from fluorescence intensity. The ethanol mass fraction ($w_{eth}$) is 0 at $x=0$ defined as the closest point to the side channel where (dye-free) water is supplied. The increase in dye concentration close to $x=0$ results in high fluorescence intensity, corresponding to a positive ethanol composition gradient. For ethanol gradient in the configuration of three zones in Figure \ref{acquisition}A, several assumptions are made: (a) At the boundary between zone 2 and 3, the ethanol mass fraction on the side of zone 3 is assumed to be equal to the composition of the octanol-rich subphase immediately after being formed by phase separation ($w_{eth23,3}$). (b) At the boundary of zone 1 and 2, the ethanol composition on the side of zone 2 is assumed to be the same as the saturated water-rich subphase from phase separation ($w_{eth12,2}$). (c) From the boundary between zones 2 and 3, the ethanol fraction in the liquid in zone 2 ($w_{eth23,2}$) is proportional to the fluorescence intensity ($I_{12,2}$) and composition from zone 2 at the boundary location between 1 and 2 ($x_{eth12,2}$). For 30\% octanol in solution A, $w_{eth23,3}$ is estimated to be 0.375 and $w_{eth12,2}$ to be 0.326, based on UNIFAC equilibrium calculations \cite{Arends2021}. Linear interpolation for assumption (c) yields an expression for the ethanol mass fraction as a function of fluorescence intensity: 
\begin{equation}
w_{eth23,2} = w_{eth12,2} \times I_{23,2}/I_{12,2}
\end{equation}
\noindent

Composition gradients were evaluated in the boundary region, in a coarse-grained manner, as illustrated in Figure \ref{acquisition}A, and in a more fine-grained manner, within and adjacent to key features on a boundary, as illustrated in Figure \ref{acquisition}B. Coarse-grained gradients were obtained from fluorescence intensity profiles of the ethanol mass fraction ($w_{eth}$) along the x-direction ($x$) at a fixed time and fixed y values. Tables of intensity values ($I(x)|_{y,t}$) were normalized using the intensity at the side channel position ($I_{x0}$) as $I(x)/I_{x0}$. The normalized intensity data contained significant image-related noise, which was reduced using a Savitzky-Golay filter in MATLAB.

Fine-grained gradients, used to evaluate the effect of diffusive boundary shape on the mobility of droplets, were obtained by taking the intensity profile along fixed lines perpendicular to local features of the diffusive boundary between solution A and B. The data acquisition process is illustrated in Figure \ref{acquisition}B, where the boundary has a triangular feature. The normalized intensity profile ($I_{norm}$), as a function of position ($d$) across the boundary region, was fit to a Gaussian decay model:

\begin{equation}
I_{norm}(d) = \frac{I(d)-I_{df}}{I_{di}-I_{df}} \sim \exp (-\frac{d^2}{a(x)|_{t}})
\label{gradinprotrusion}
\end{equation}

\noindent
where $I(d)$, $I_{di}$, and $I_{df}$ are the mean gray values at position $d$ along the diffusive boundary, the initial position at the center of the mixing zone and the final position at the water-rich side, respectively. The model parameter $a(x)|_{t}$, was fitted to $I_{norm}(d)$ profiles using least squares regression. $a(x)|_{t}$ values were used to illustrate the difference in sharpness of intensity profiles with x-position at a fixed time. Normalized intensity values were converted to ethanol mass fraction profiles with the boundary conditions of the water-rich phase $w_{eth}(I_{df})$ and the initial solution A composition $w_{eth}(I_{di})$.

\section{Results and Discussion}

\subsection{Speed and direction of model oil-rich microdroplets mediated by local composition gradients}

\begin{figure}[H]
	\includegraphics[trim={0cm 0cm 0cm 0cm}, clip, width=0.85\columnwidth]{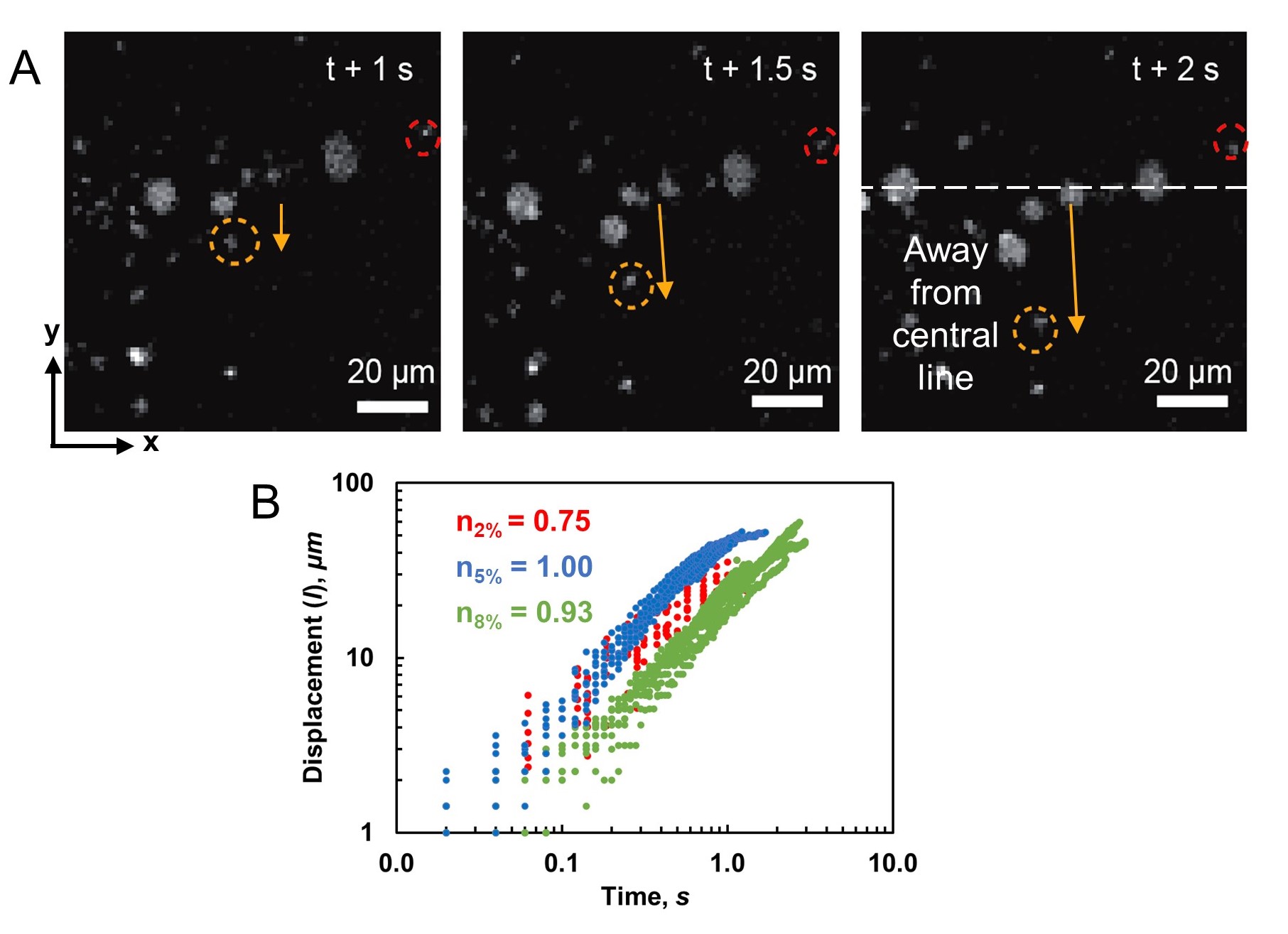}
	\caption{(A) Time-based optical images of droplets moving away from a triangular protrusion for solution A with 8 wt-\% octanol. The red dotted circle shows a stationary object as a reference. A representative moving droplet is indicated with a yellow dotted circle. (B) Octanol-rich droplet displacement versus time from a stable triangular protrusion for solution A with 2 wt-\% (red), 5 wt-\% (blue) and 8 wt-\% (green) octanol.
	}
	\label{regime4}
\end{figure}

At low octanol mass fraction in solution A from 2 to 20 wt-\% (Composition 3 to 8 in Table \ref{Experimental Compositions Table}), a trail of octanol-rich microdroplets is produced. Microdroplets form and are propelled away from the diffusive boundary between water and solution A. The locations of droplet formation are usually at the tip of a triangular protrusion of solution A into the side of the penetrating water. The reason for localized droplet formation is that the composition gradient of octanol is the sharpest at these locations \cite{Zhang2020}. 
Figure \ref{regime4}A shows the trajectory of a representative droplet after formation. The droplet displacement increases with time with an exponent of $0.75<n<1$. Typically, movement is away from a central line drawn through a protrusion, oblique to the x-axis. There does not appear to be a strong dependence on the initial octanol concentration at low octanol concentrations, as shown in Figure \ref{regime4}B. 

\begin{figure}[H]
	\includegraphics[trim={0cm 0cm 0cm 0cm}, clip, width=1\columnwidth]{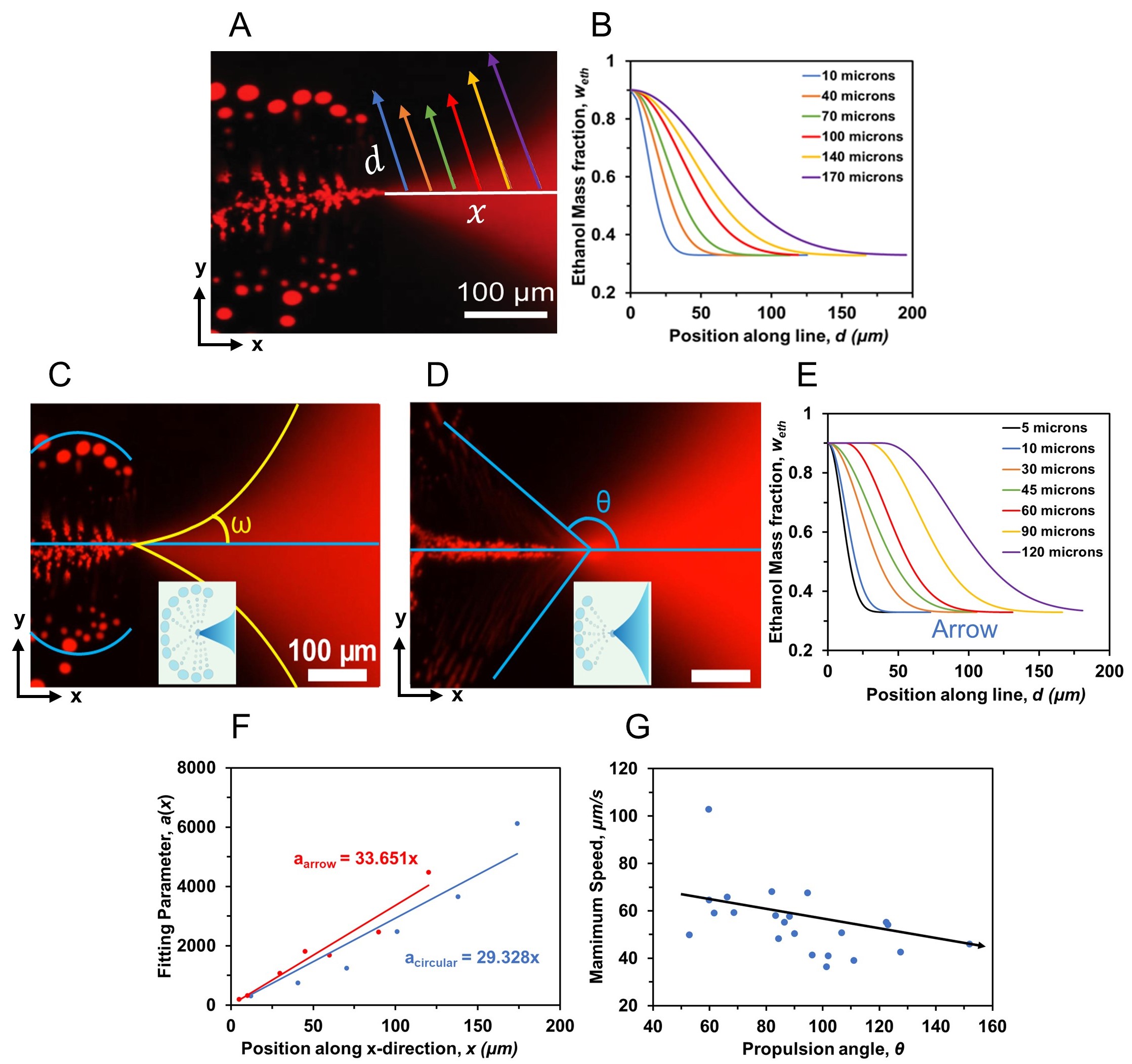}
	\caption{Effect of composition gradients on droplet movement pattern for solution A with 10 wt-\% butylparaben. (A) Fluorescence image of typical a triangular protrusion annotated with data acquisition lines for (B). (B) Ethanol fraction as a function of position along the lines shown in (A). Fluorescence image of (C) semi-circular ($\omega = 14 \degree$) and (D) arrow-shaped ($\omega = 27 \degree$) droplet movement patterns. $\theta$ and $\omega$ are defined as the angle with the x-axis. (E) Ethanol fraction as a function of position along an arrow-shaped triangular zone. (F) Parameter $a$, defined in Equation \ref{gradinprotrusion} values as a function of x-position for semi-circular and arrow shapes. (G) Maximum droplet speed as a function of propulsion angle of droplets $\theta$ from phase separation zone.}
	\label{nozzleconcentrationgradient}
\end{figure}

We expected the shape of the protrusion to impact the direction and speed of the microdroplets and illustrate this effect using the ternary mixture of butylparaben, ethanol and water. The fluorescence images in Figure \ref{nozzleconcentrationgradient} clearly show a protrusion arising in solution A with 10 wt-\% butylparaben in ethanol and the trails of stationary and mobile butylparaben-rich droplets immediately after their formation. The local ethanol mass fraction profile was obtained from the fluorescence intensities along to the protrusion and is shown in Figure \ref{nozzleconcentrationgradient}B. The fluorescence intensity is high at the point of butylparaben-rich droplet formation and decreases in the negative x-direction towards the water side. The ethanol composition gradient is sharpest in the butylparaben-rich continuous phase near the point of formation. 

Figures \ref{nozzleconcentrationgradient}C and D show two distinct patterns of droplet movement from the phase separation zone. The first pattern is semi-circular with an opening angle of $\omega$ = 14$\degree$. The second pattern shown in Figure \ref{nozzleconcentrationgradient}D possesses an arrow shape with an angle of $\omega$ = 27$\degree$. Figure \ref{nozzleconcentrationgradient}F shows the difference in composition gradients for these two droplet movement patterns. The parameter $a$ at fixed $t$ and $x$, defined in Equation \ref{gradinprotrusion}, has lower values for steeper x-direction composition gradients. Droplet movement in semi-circular patterns experiences sharper composition gradients along the entire triangular zone compared to the arrow pattern.

Figure \ref{nozzleconcentrationgradient}G shows the relation between maximum droplet speed and the angle from the x-axis. While the data are scattered, maximum droplet speed decreases as the angle increases. Droplets moving at higher angles encounter smaller composition gradients and hence less interfacial stress, which reduces the maximum speeds that the droplets can achieve compared to droplets moving at smaller angles.

\subsection{Composition of microdroplets in spatially segregated zones arising from liquid-liquid phase separation}

\begin{figure}[H]
	\includegraphics[trim={0cm 0cm 0cm 0cm}, clip, width=0.7\columnwidth]{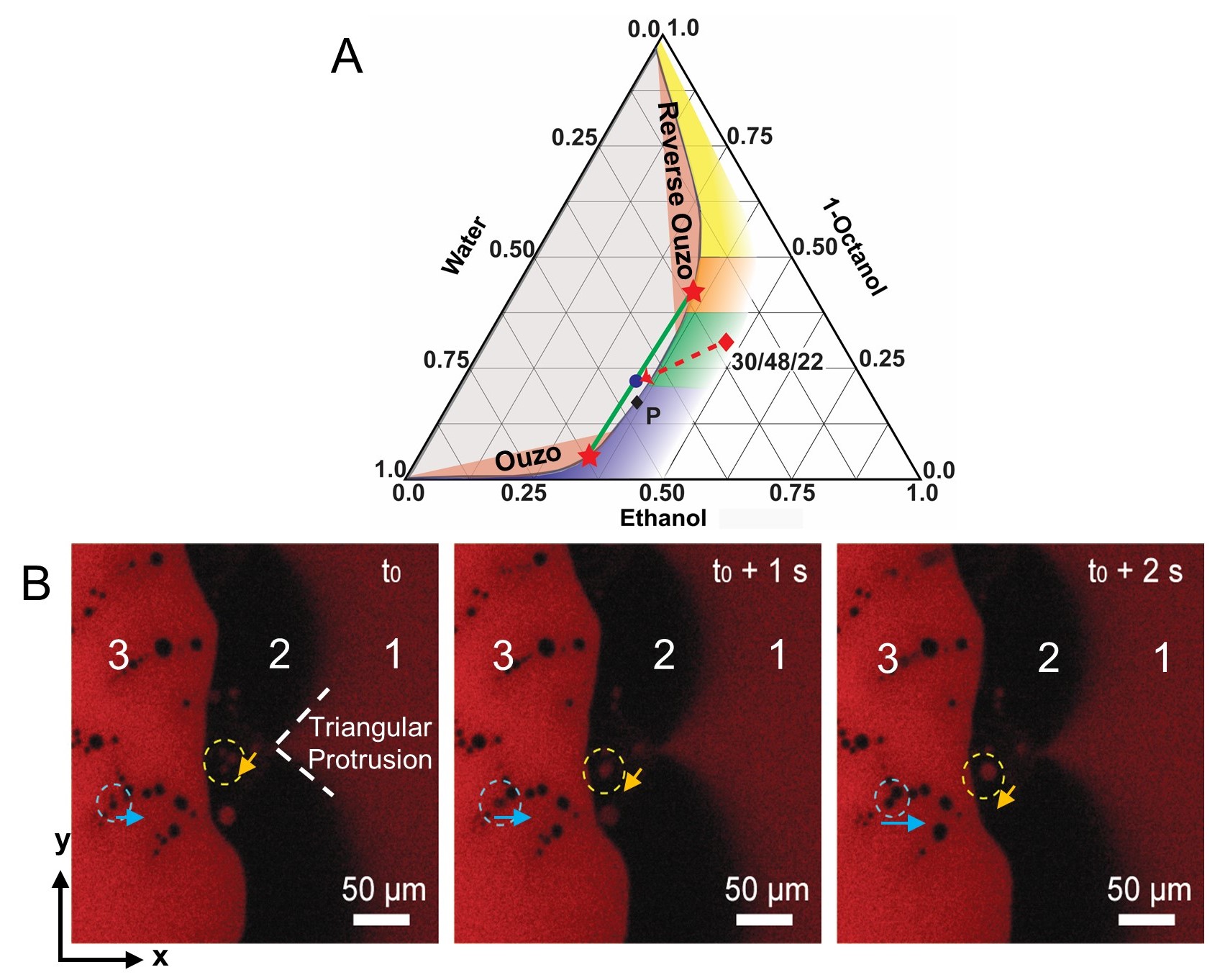}
	\caption{(A) Ternary phase diagram indicating the composition of solution A and of the subphases formed from phase separation. Compositions are in mass fractions. Red diamond: solution A; Black diamond: plait point; Stars: subphases; Green line: tie line; Red dotted line: composition path before phase separation; Pink areas: ouzo region (octanol droplets in water-rich subphase) and reverse ouzo region (water droplets in octanol-rich subphase). Reproduced from \textit{Energy Fuels} \textbf{2021}, 35, 6, 5194–5205 (\cite{Arends2021}). Copyright 2021 American Chemical Society. (B) A time sequence showing magnified fluorescence images of the three zones developed from phase separation. Zone 1: solution A; Zone 2: water-rich subphase with octanol-rich microdroplets (indicated by orange dashed circle); Zone 3: octanol-rich phase with water-rich droplets (indicated by a blue dashed circle). The orange and blue arrows show directions of droplet movement.
	}
	\label{overviewregime3}
\end{figure}

Mixing of water with solution A containing octanol from 30 to 35 wt-\% produces both oil-rich and water-rich microdroplets, in addition to three spatially separated zones. If we focus on the behaviour for 30 wt-\% octanol in solution A (composition 2 in Table \ref{Experimental Compositions Table} and annotated in the phase diagram in Figure \ref{overviewregime3}A), three well-defined zones are evident in the images (Figure \ref{overviewregime3}B). The lipophilic dye indicates the spatial distribution of oil-rich phase domains. A continuous octanol-rich subphase is labelled as zone 3 (close to the side channel) that extends along positive x-direction. A continuous water-rich subphase is labelled as zone 2 (further from the side channel) that separates zone 1 and 3. Solution A is labelled as zone 1 (furthest from the side channel) that recedes along the positive x-direction. From the phase diagram, tie lines between the water-rich and octanol-rich subphases intersect with ouzo and reverse ouzo regions. Zone 2 and 3 form from primary liquid-liquid phase separation by diffusive mixing water with solution A. The water-rich microdroplets form in zone 3 after the primary phase separation already creates spatially segregated zones.

\subsection{The motion of oil-rich microdroplets in a water-rich zone}

\begin{figure}[H]
	\includegraphics[trim={0cm 0cm 0cm 0cm}, clip, width=0.95\columnwidth]{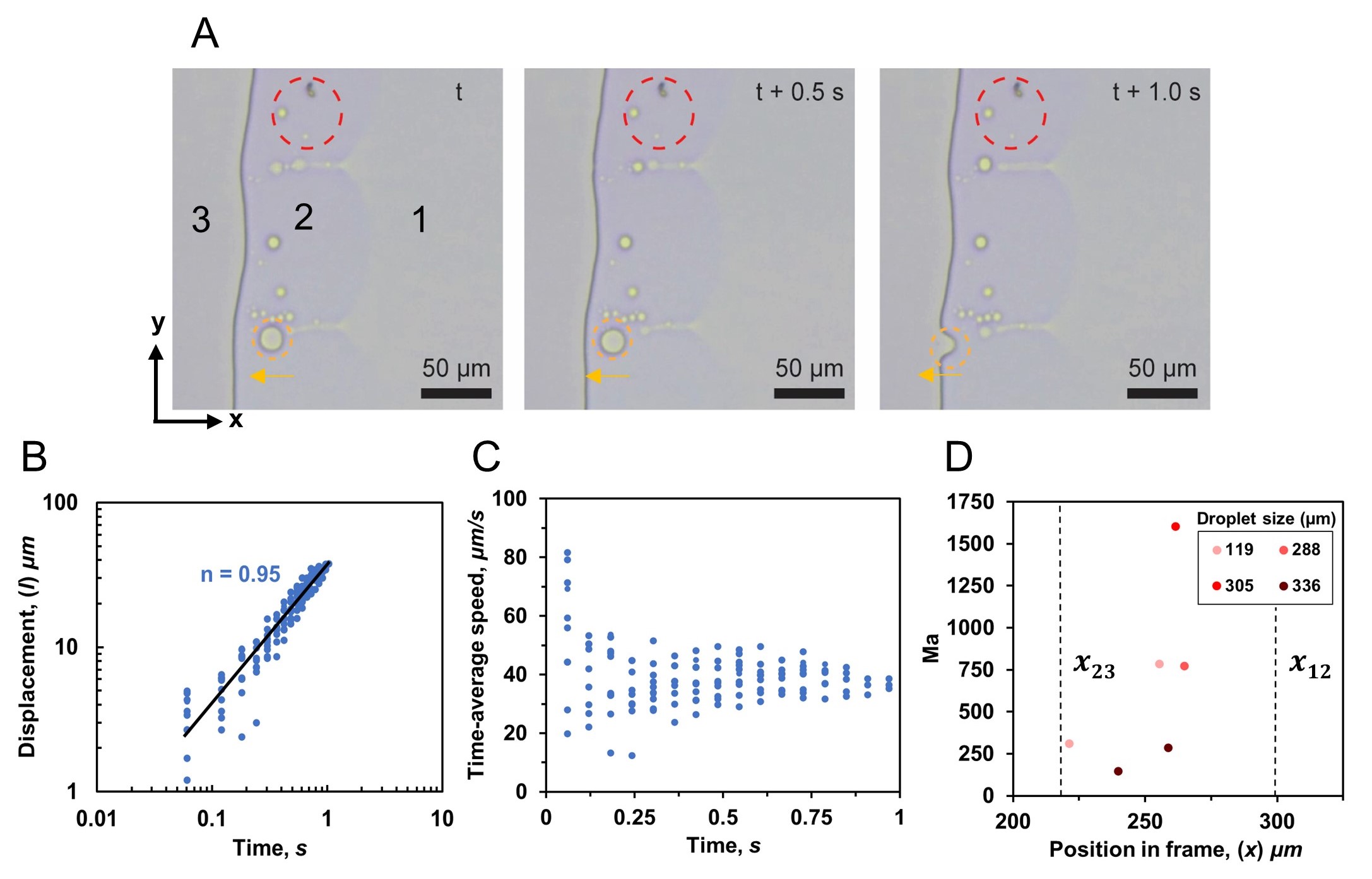}
	\caption{Oil-rich droplet displacement for solution A with 30 wt-\% octanol. (A) Time-lapsed optical images of octanol-rich droplets formed from triangular protrusions along the zone 2 boundary. The red dashed circle indicates a reference object, and the yellow dashed circle shows a representative octanol-rich droplet. The three zones visible in the field of view are labelled. (B) Displacement of octanol-rich droplets with time after release. (C) Time-averaged octanol-rich droplet speed as a function of time. (D) Marangoni number of octanol-rich droplets as a function of x-direction position at four y-direction positions and at fixed $t_0$. Triangular protrusions are located at y = 106 ${\mu}m$ and y = 312 ${\mu}m$. ($x{12}$ and $x_{23}$) indicate the boundary between zone 1 and 2 and the boundary between zone 2 and 3.
	}
	\label{regime3oilrich}
\end{figure}

Figure \ref{regime3oilrich}A shows the movement of one representative octanol-rich microdroplet in zone 2. These droplets are typically 3 ${\mu}$m to 9 ${\mu}$m in lateral diameter, so the channel is only 3 to 4 times larger than the dimension of the largest microdroplets present. The nucleation of octanol-rich droplets in water-rich zone 2 occurs from triangular-shaped protrusions extending from zone 1 (solution A). At the time $t$ (Figure \ref{regime3oilrich}A), an octanol-rich droplet detached from the triangular protrusion at the boundary between zone 1 and 2 and propelled in the negative x-direction until it merged with the octanol-rich subphase at the boundary of zone 2 and 3 after 1 $s$. 

Figure \ref{regime3oilrich}B shows the displacement of ten individual octanol-rich droplets in zone 2 as a function of time, which scales linearly with time ($l \sim t^{0.95}$). The octanol-rich droplets have a broad initial displacement rate which converges to an average displacement rate of (41.9 $\pm$ 4.3) ${\mu} m/s$ near the zone 2 to zone 3 boundary (Figure \ref{regime3oilrich}C). The initial and average speeds of octanol-rich droplets in zone 2 are notably slower than for water-rich droplets in zone 3, as will be shown in Figure \ref{regime3waterrich}. 

\begin{figure}[H]
	\includegraphics[trim={0cm 0cm 0cm 0cm}, clip, width=0.80\columnwidth]{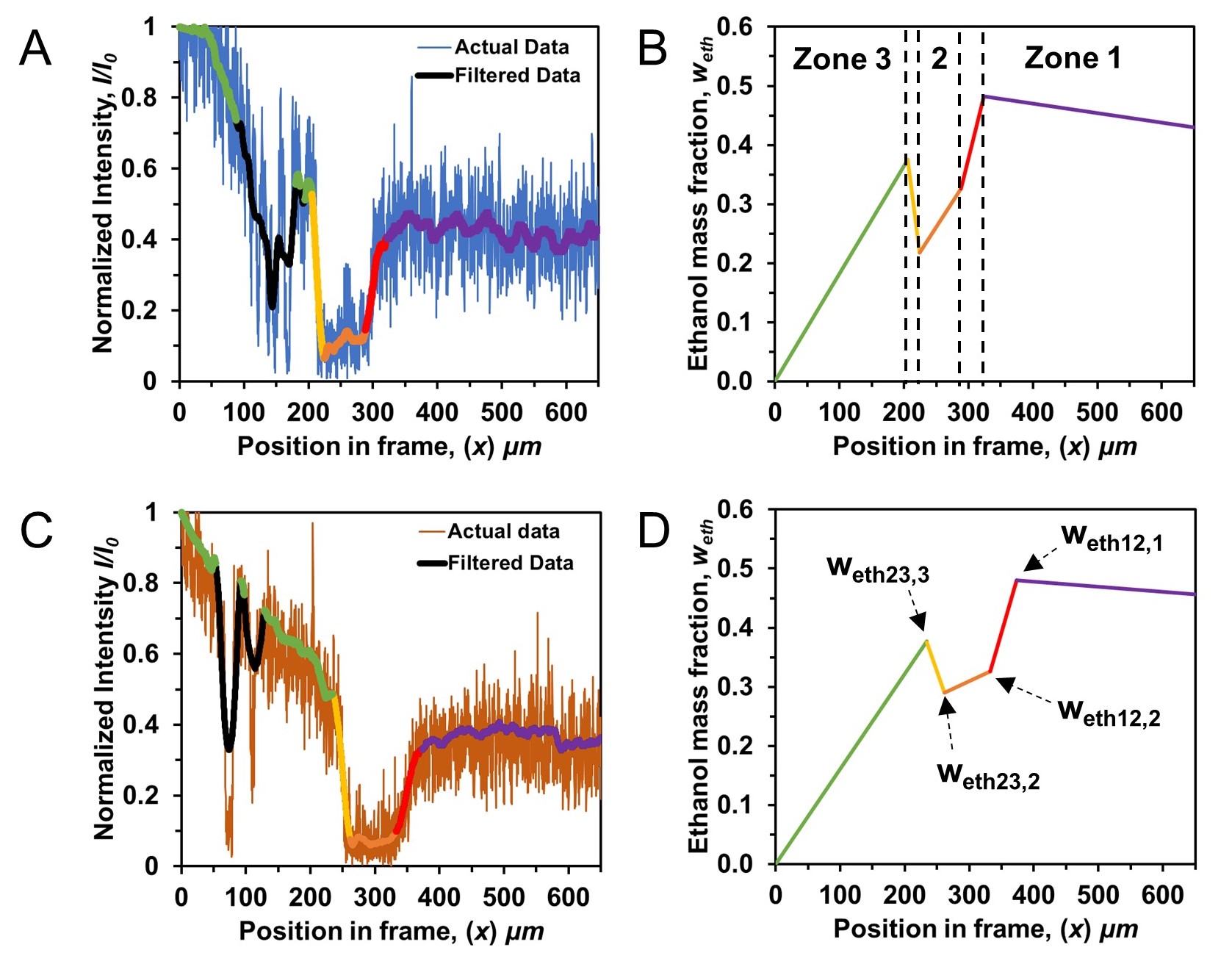}
	\caption{Concentration gradients as indicated by fluorescence intensity at $t_0$ for 30 wt-\% octanol in solution A. (A) and (C) Normalized intensity profile as a function of x-position. (B) and (D) Ethanol mass fraction as a function of x-position calculated from intensity profiles. Different colours represent the slope along each zone and boundary.
	}
	\label{regime3concentration}
\end{figure}

Figures \ref{regime3concentration}A and C show intensity profiles at two y-values once they have been normalized with respect to the intensity signal from dye-free water at the channel entrance ($x=0$). Although the plots include image-related noise, the general trend is clear. The fluorescence intensity is almost constant in zone 1. It drops sharply to a valley in zone 2 and increases gradually in zone 3.  

Spatial gradients in ethanol mass fraction within zone 2 are evaluated from fluorescence intensity (\ref{regime3concentration}A). The ethanol mass fraction gradient in zone 2 is (1.99 $\pm$ 1.26) $\times 10^{-3}$ ${\mu m^{-1}}$ in the positive x-direction. The ethanol mass fraction gradient jumps to ($4.02 \pm 1.19) \times 10^{-3}$ ${\mu m^{-1}}$ in the positive x-direction at the boundary between zone 1 and zone 2. The change in gradient at this boundary is the sharpest at or near a triangular protrusion location and becomes smaller at locations remote from triangular protrusion. The center of such protrusions is the location where phase separation occurs, causing the sharp concentration gradient. 

The gradient of the ethanol mass fraction in zone 2 has important implications for oil-rich microdroplet motion. The interfacial tension between an oil-rich microdroplet and the water-rich continuous phase decreases with increasing ethanol mass fraction. The interfacial tension gradient along a droplet surface in the negative x-direction induces Marangoni stress so that the oil-rich microdroplets move to the direction with higher interfacial tension, consistent with what has recently been reported \cite{Zhang2020}.

The composition gradient applicable to octanol-rich droplets can also be demonstrated with the dimensionless Marangoni number. In this experiment (experiment 2), the composition gradient across an octanol-rich droplet was obtained from ethanol mass fraction as a function of position in zone 2.  The Marangoni number values range from $150<Ma<1600$ with x-direction (Figure \ref{regime3oilrich}D) and do not appear to be a function of the y-position value. Here $x=0$ is defined at the position of the side channel. The Marangoni values are higher than the case of a sub-millimetre oil drop immersed in a stratified ethanol/water liquid ($Ma<10^{3}$)\cite{li2021marangoni} but lower than a water/glycerol evaporating drop ($Ma<10^{5}$) \cite{diddens2021competing}, where both thermal and composition gradients enhance Marangoni flows. The large $Ma$ number in our experiments may be due to the large surface tension gradients $\Delta \sigma $ arising from the local composition gradients created by liquid-liquid phase separation in confined space.

The effect of the triangular protrusion on droplet displacement is also observed here, similar to Figure \ref{nozzleconcentrationgradient}. Droplets closer to the triangular protrusion in both x and y have higher Marangoni numbers related to sharper composition gradients near the protrusions. The effect in the y-direction is evinced by the high Marangoni number of the droplet at (x = 261, y = 288), which had the smallest distance from a triangular protrusion in the data set. 

\begin{figure}[H]
	\includegraphics[trim={0cm 0cm 0cm 0cm}, clip, width=0.85\columnwidth]{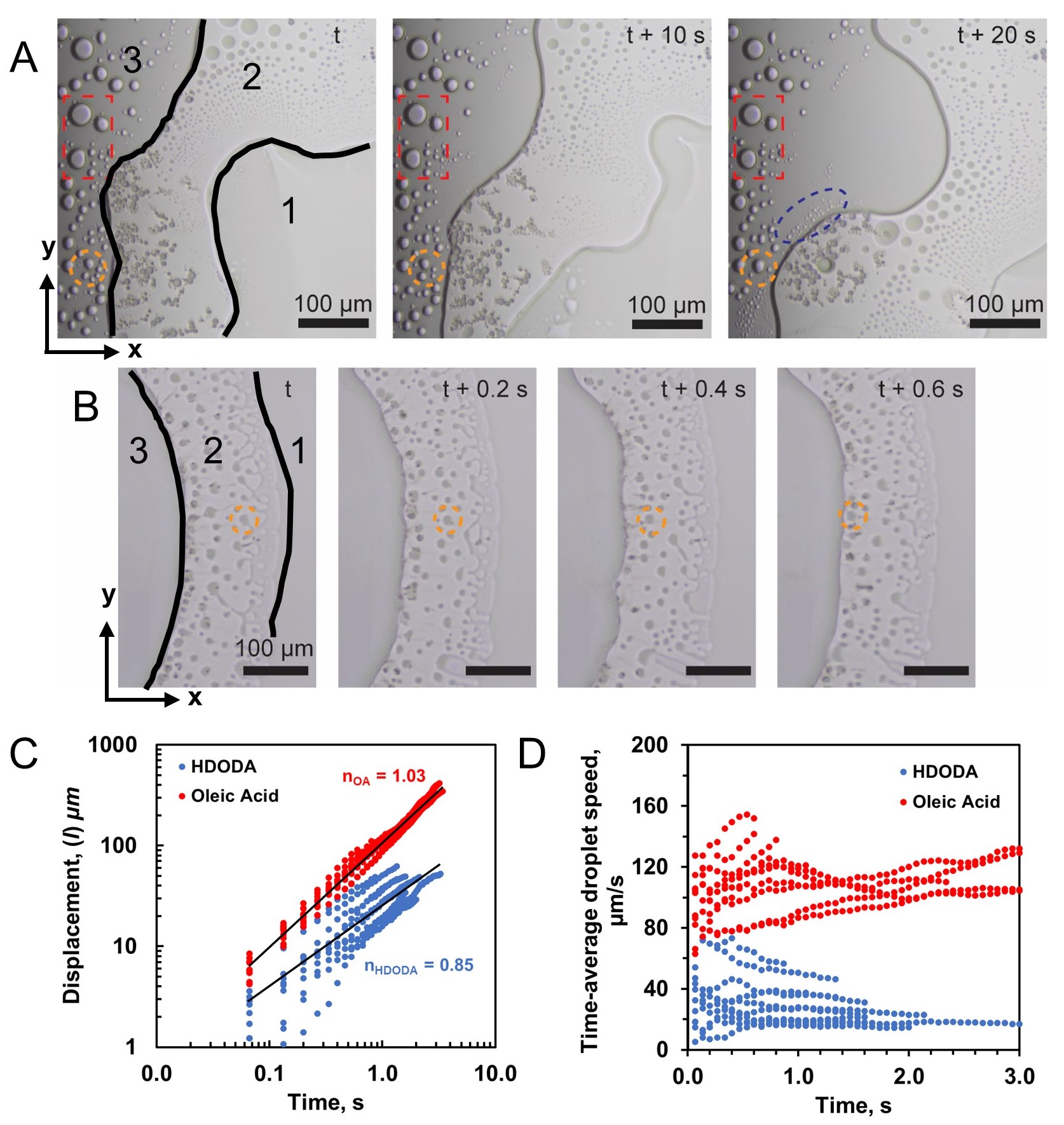}
	\caption{Time-based optical images of three-zone configurations for model oils HDODA and Oleic acid in solution A. (A) 25 wt-\% HDODA/50 wt-\% ethanol/25 wt-\% water in solution A. The red dashed box indicates reference stationary water-rich droplets within the oil-rich zone. The yellow dashed circle indicates a representative water-rich droplet that coalesces with nearby growing droplets. The blue dashed circle indicates the spontaneous growth of small water-rich droplets close to the moving boundary. (B) 24 wt-\% Oleic Acid/49 wt-\% ethanol/27 wt-\% water as solution A. Yellow dashed circle indicates the movement of one representative oleic acid-rich droplet towards the oil-rich zone with time. (C) Displacement of oil-rich droplets with time for HDODA (blue) and oleic acid (red) as model oils. Linear fit applied to show scaling. (D) Time-averaged oil-rich droplet speed as a function of time.
	}
	\label{HDODA OA}
\end{figure}

Two additional ternary systems were examined to further demonstrate the features of propelling oil-rich microdroplets and the evolution of the oil-rich zone. As shown in Figure \ref{HDODA OA}, a similar three-zone configuration is formed for solution A comprising HDODA or oleic acid, water and ethanol as compositions labelled in the phase diagrams shown in Figure \ref{ternarydiagrams}. Lines of HDODA-rich droplets are ejected from the undulating boundary between the water-rich zone and solution A. These droplets coalesce rapidly and merge with the HDODA-rich zone on the other side of the water-rich zone. The speed of HDODA-rich microdroplets ranges from 20 to 80 $\mu m/s$, slower than octanol-rich microdroplets in general. The process reveals that the oil-rich zone develops from merged propelling microdroplets. Meanwhile, the water-rich droplets within the HDODA-rich zone nucleate and grow on the surface, remaining immobilized throughout the entire process.   

By contrast, no water-rich microdroplets are formed in the oil-rich zone for the oleic acid, ethanol and water ternary. The boundary between the water-rich zone and the oleic acid solution is rough and evolves rapidly with time. The water-rich subphase forms an elongated zone. Microdroplets of oleic acid form and pinch off along the entire boundary instead of at the tips of protrusions. The speed of oleic acid-rich microdroplets can reach 150 $\mu m/s$, faster than that of HDODA-rich and octanol-rich microdroplets. In addition, the droplets travel for a longer distance without coalescing with nearby droplets. The results for the diverse mixtures demonstrate that the propelling droplet behaviour and development of segregated zones are general phenomena for liquid-liquid phase separation in confined spaces, although the speed of droplets and zone boundary shape may vary among ternary solutions. 

\subsection{The motion of water-rich microdroplets in an oil-rich zone}

At low oil concentration in solution A, the fast motion of propelling oil-rich microdroplets induces a directional flow in the narrow chamber to replenish the liquid loss at the location of phase separation due to mass conservation. Such directional transport induced by propelling oil-rich microdroplets is also expected to be present at high oil concentrations with separated zones. Without tracer nanodroplets at high oil concentration, the presence of the induced flow can actually be inferred from the motion of water-rich microdroplets. 

\begin{figure}[H]
	\includegraphics[trim={0cm 0cm 0cm 0cm}, clip, width=1\columnwidth]{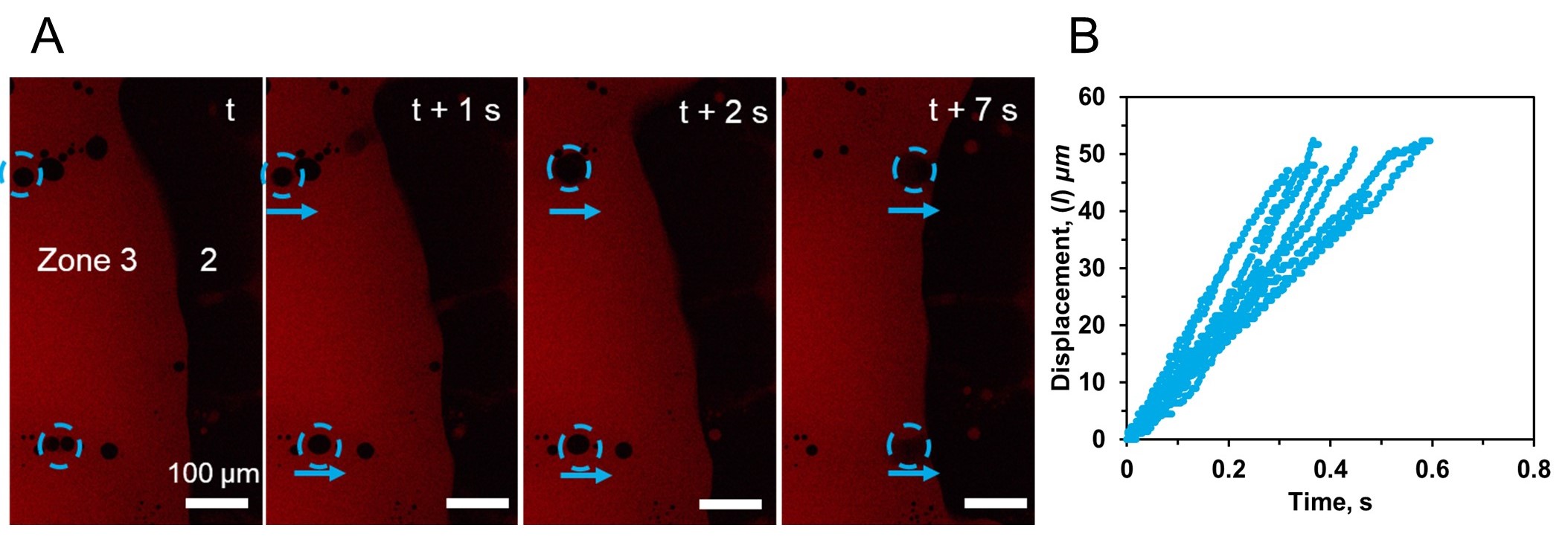}
	\caption{(A) Time-based fluorescence image of water-rich droplet movement at the 30 wt-\% octanol composition. Dashed circles indicate the movement of two water-rich droplets within zone 3. (B) Displacement of water-rich droplets in zone 3 as a function of time. Each set of dots represents the motion of one droplet measured with a high-speed camera. The final displacement position of each droplet is when they merge with the boundary between zone 2 and 3.
	}
	\label{regime330water}
\end{figure}

Behind the protrusion in zone 2, water-rich microdroplets form on the surface in zone 3. These droplets give rise to low fluorescence intensity signals within a high fluorescence intensity phase, suggesting the droplets consist mainly of a water-rich solution (Figure \ref{regime330water}A). Unlike oil-rich microdroplets, water-rich droplets do not propel immediately after their formation. Near the oil-rich zone 3 boundary, some water-rich microdroplets detach from the surface and move suddenly in the positive x-direction. After a short time, these moving water-rich microdroplets merge with zone 2. The displacement ($l$) of representative water-rich droplets show a clear linear relationship with time ($t$), as illustrated in Figure \ref{regime330water}B. The linearity of their motion indicates that this droplet movement cannot be attributed to Brownian motion where $l$ is expected to scale with $t^{1/2}$. 

The motion of those water-rich microdroplets is not due to Marangoni stress in zone 3. The fluorescence intensity decreases in the x-direction in zone 3, as shown in Figure \ref{regime3concentration}, suggesting that the ethanol mass fraction in the continuous phase increases as the boundary between zone 3 and 2 is approached. The highest ethanol mass fraction is located at the boundary where the ethanol mass fraction in zone 3 has not been influenced by water diffusion as the oil-rich subphase is newly formed from the phase separation. The ethanol mass fraction in the oil-rich continuous phase is estimated from the fluorescence intensity, as outlined in the Experimental Section. A spatial gradient of ($1.72 \pm 0.07) \times 10^{-3}$ ${\mu m^{-1}}$ is obtained from six y-position values in zone 3. 

In zone 3, the majority of water-rich microdroplets far from the boundary of the two zones remain stationary. The stability of these water-rich microdroplets suggests that the high viscosity of the oil-rich continuous phase may hinder the movement of water-rich microdroplets, even in the presence of a large composition gradient. For those moving water-rich droplets, their motion is toward lower surface tension (i.e. higher ethanol fraction). The gradient in the ethanol fraction is expected to pull water-rich microdroplets in the negative x-direction opposite to what is observed. Instead of the Marangoni effect, the movement of these water-rich droplets is attributed to the directional flow induced by propelling oil-rich microdroplets in the water-rich zone. In other words, the motion of water-rich microdroplets in zone 3 originates from the composition gradient in zone 2 through induced flow. 

\begin{figure}[H]
	\includegraphics[trim={0cm 0cm 0cm 0cm}, clip, width=1\columnwidth]{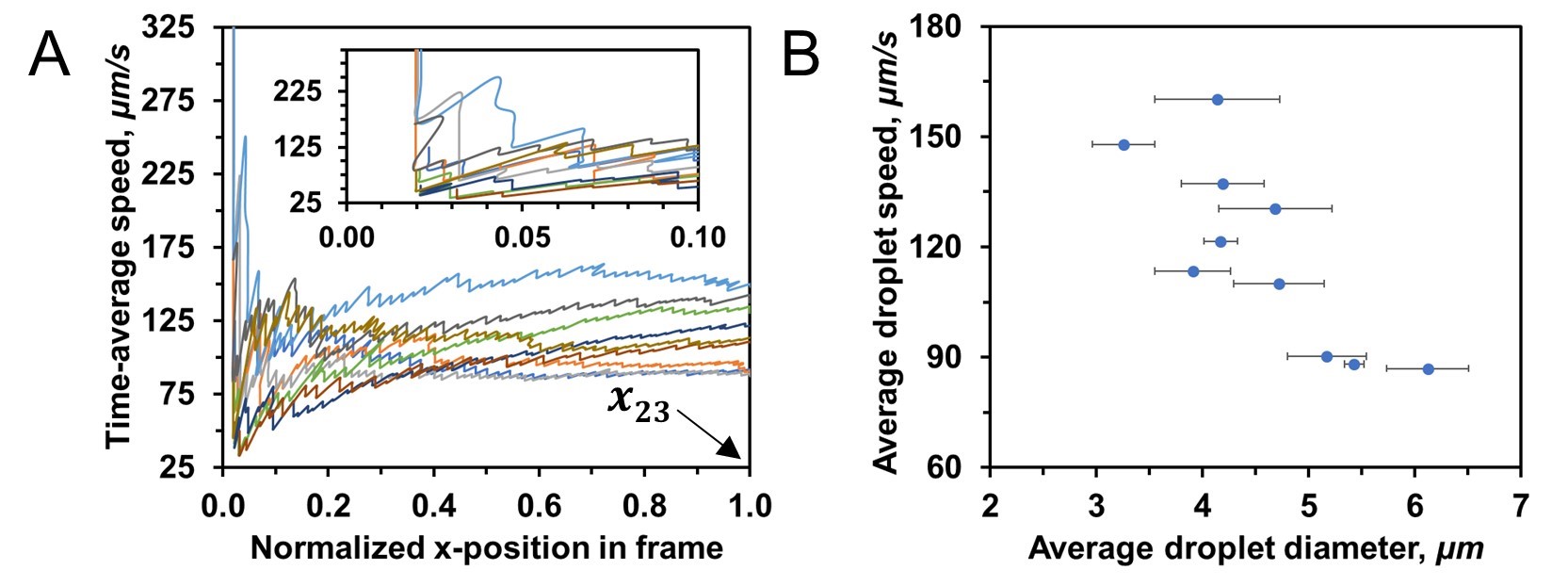}
	\caption{(A) Time-averaged droplet speed as a function of normalized droplet displacement. Inset: zoom-in plot of early displacement. (B) Water-rich droplet speed as a function of average droplet size at 30 wt-\% octanol composition.
	}
	\label{regime3waterrich}
\end{figure}

Figure \ref{regime3waterrich}A shows the time-averaged speed of nine water-rich droplets in zone 3 over a displacement window of (47.2 $\pm$ 10.3) $\mu$m. Their trajectories are characterized by high initial speeds that decay within a few microns to a minimum of (58 $\pm 20$) ${\mu}m/s$. Droplet speeds then gradually increase to an average speed of (119 $\pm$ 24) $\mu m/s$. Droplets merge with the boundary at $d_{final}$ (total distance travelled by the droplet to reach the boundary between zones 2 and 3 at $x_{23}$).
As the driving force from induced flow and the drag effect from the hydrophobic bottom surface and the surrounding octanol-rich subphase may reach transient balance, the speed of the water-rich droplets remains relatively constant until they reach the boundary between zones 2 and 3. 

The Reynolds number, $Re$, of the water-rich droplets is of the order $10^{-4}$ calculated with
\begin{equation}
Re=\frac{UR}{\nu}
\end{equation}
\noindent
where R is the droplet radius, $\nu$ the kinematic viscosity of the surrounding liquid in zone 3, and $U$ is the relative droplet speed \cite{Lohse2020}. For such low Reynolds numbers, viscous dissipation dominates. 

The viscosity of the oil-rich continuous phase in zone 3 is high, which may lead to faster decay in the speed of water-rich microdroplets after the initial rapid motion, as shown in the insert of Figure \ref{regime3waterrich}A. As the bottom surface is hydrophobic, this implies that the water-rich microdroplets experience a strong drag force as they displace oil-rich subphase from the surface. The shear arising from the directional flow is sufficiently strong to overcome both viscous dissipation from the oil-rich continuous phase and the drag force from the hydrophobic surface.

Figure \ref{regime3waterrich}B provides the average speed of water-rich droplets with diameters from 3 to 6.5 $\mu$m. For a given droplet diameter, the speed scatters in a large range. However, the overall trend is toward lower speeds for larger droplets. Water-rich droplets furthest from the zone 3 boundary shown in Figure \ref{regime3concentration}C, near ($x=0$), are typically larger in size due to coalescence with nearby droplets. Some of the large water-rich droplets remain immobilized on the hydrophobic bottom surface, suggesting that the experienced drag forces must be overcome to trigger droplet motion. Microdroplets closer to the zone 3 boundary are smaller in size and are more mobile due to a reduced surface area. 
 
\begin{figure}[H]
	\includegraphics[trim={0cm 0cm 0cm 0cm}, clip, width=1\columnwidth]{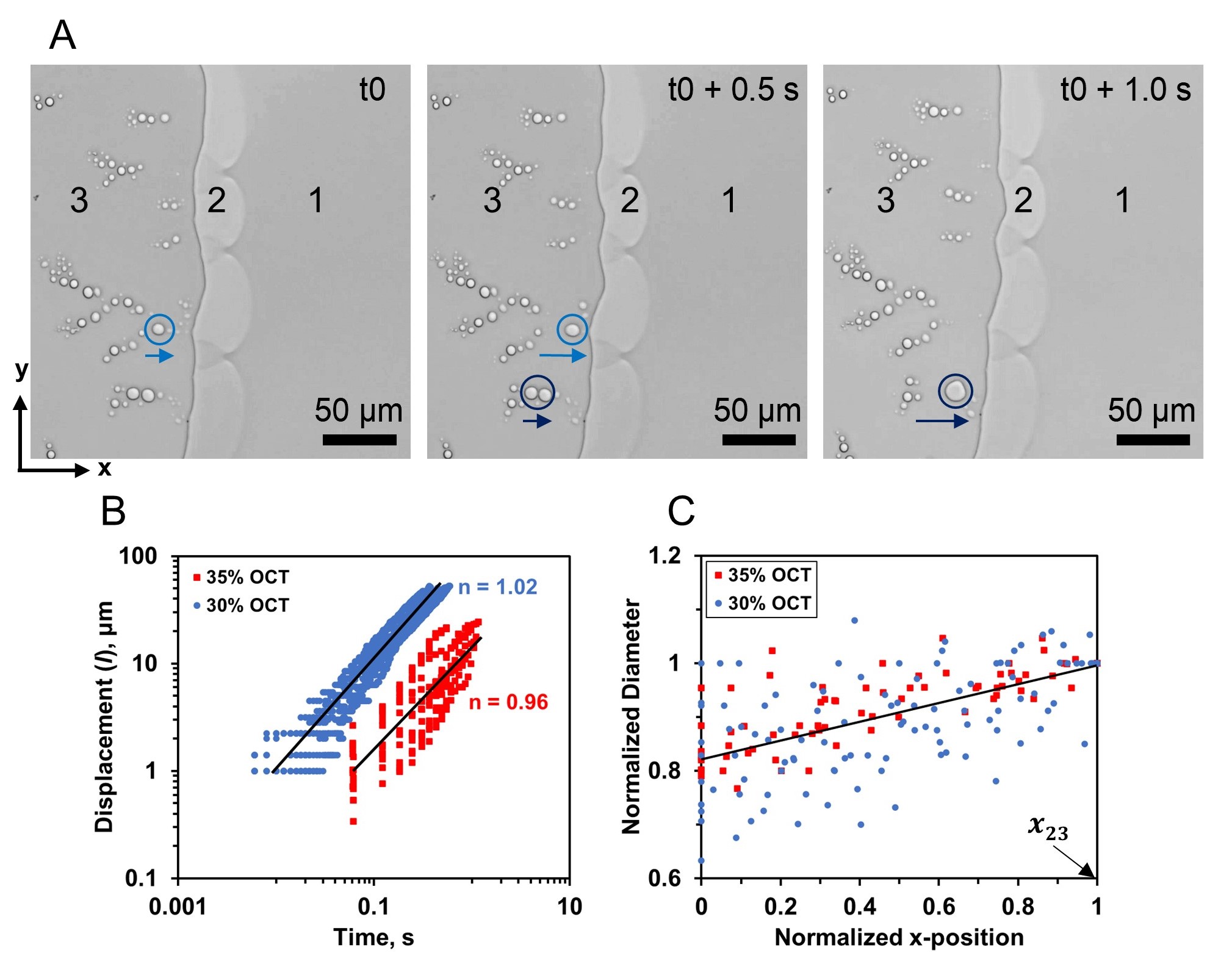}
	\caption{Comparison between initial solution compositions. (A) Time-based optical images of water-rich droplet movement in zone 3 with 35 wt-\% octanol in solution A. (B) Displacement of water-rich droplets in zone 3 as a function of time with 30 wt-\% and 35 wt-\% octanol in solution A. (C) Normalized water-rich droplet diameter as a function of normalized x-position in zone 3.
	}
	\label{regime335water}
\end{figure}

For solution A with 35 wt-\% octanol, phase separation and the droplet motion are qualitatively similar to what was described above at 30 wt-\% octanol. Figure \ref{regime335water}A shows bright-field images of water-rich droplets for this case. Water-rich droplets in zone 3, far from the boundary with zone 2, form branch-like patterns observed previously for the diffusion-dominated ouzo effect in confinement \cite{Lu2017}. 

Figure \ref{regime335water}B compares the displacement of water-rich droplets with time for 30 and 35 wt-\% octanol in solution A. The water-rich droplets for solution A with 35 wt-\% octanol are displaced a shorter distance and at a slower displacement rate on average (23 $\pm$ 10 $\mu$m/s) than for solution A with 30 wt-\% octanol. The water-rich droplets show dynamic behaviour similar to the nanodroplet tracers in the replenishing flow \cite{Zhang2020}. The resulting transport mechanism differs significantly from diffusion-based transport, which scales with $l \sim t^{1/2}$ in 2D confinement \cite{Lu2017}. 

Figure \ref{regime335water}C shows the normalized diameter of water-rich mobile droplets as a function of normalized x-position, where the normalization was performed by dividing local droplet diameter by the initial droplet diameter, and x-position was normalized by subtracting the initial position value from the position value and then dividing by the distance from the initial position to the boundary between zone 2 and 3 ($x_{23}$). The mean droplet size increases as the boundary between zone 2 and 3 is approached, suggesting the water-rich phase increases within the droplet by diffusion as the droplets approach water-rich zone 2. 

\begin{figure}[H]
	\includegraphics[trim={0cm 0cm 0cm 0cm}, clip, width=1\columnwidth]{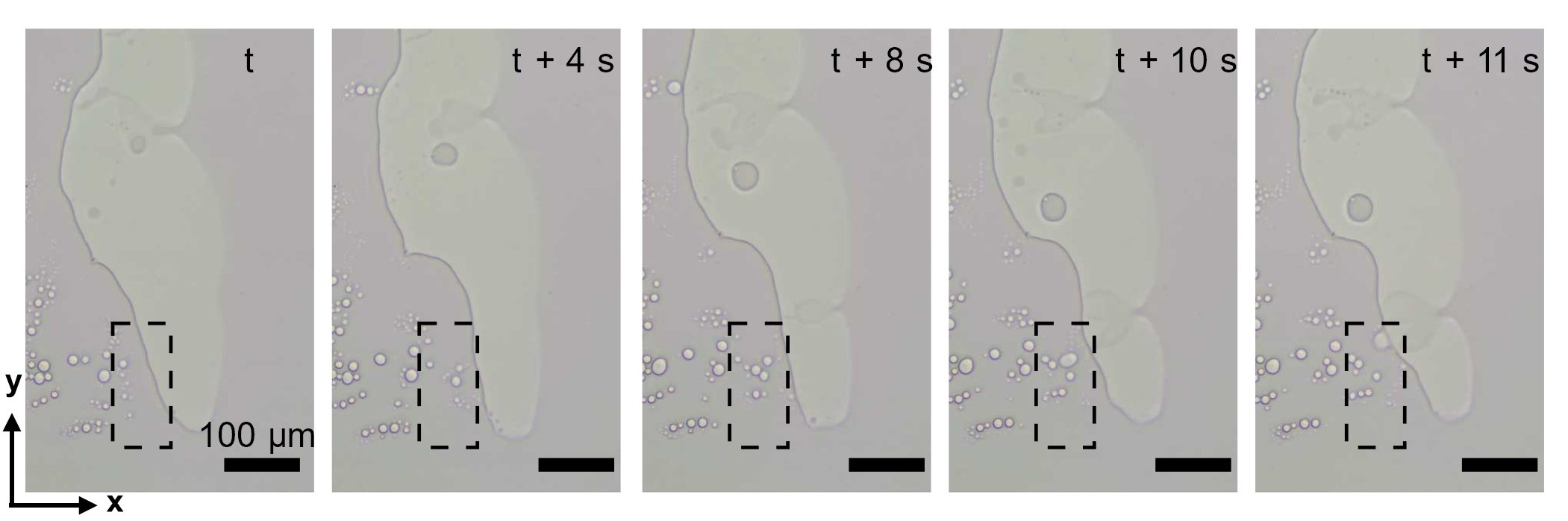}
	\caption{Time-based optical images at the 35 wt-\% octanol solution composition showing water-rich droplet nucleation, growth and subsequent coalescence.
	}
	\label{growth and movement}
\end{figure}

The motion of the microdroplets can be enhanced by coalescence, as shown in Figure \ref{growth and movement}. The images show a section of the oil-rich zone that initially did not contain water-rich droplets. With time, droplet nucleation and subsequent growth occurred from $t to t + 8 s$. At $t + 10 s$, two droplets coalesce, driving the motion of the now larger droplet toward the boundary with the water-rich zone. The phenomenon of a jumping drop induced by coalescence of two parent drops has been investigated extensively \cite{liu2021coalescence, wang2017critical,chen2007dropwise,chen2018numerical,lv2013condensation}. The reduction in the interfacial energy from coalescence contributes to the kinetic energy of the jumping drop. It is notable that in the confined space, droplet coalescence can also enhance the motion of the merged drop. 

\subsection{A comparison of the speed of octanol-rich and water-rich microdroplets}

\begin{figure}[H]
	\includegraphics[trim={0cm 0cm 0cm 0cm}, clip, width=1\columnwidth]{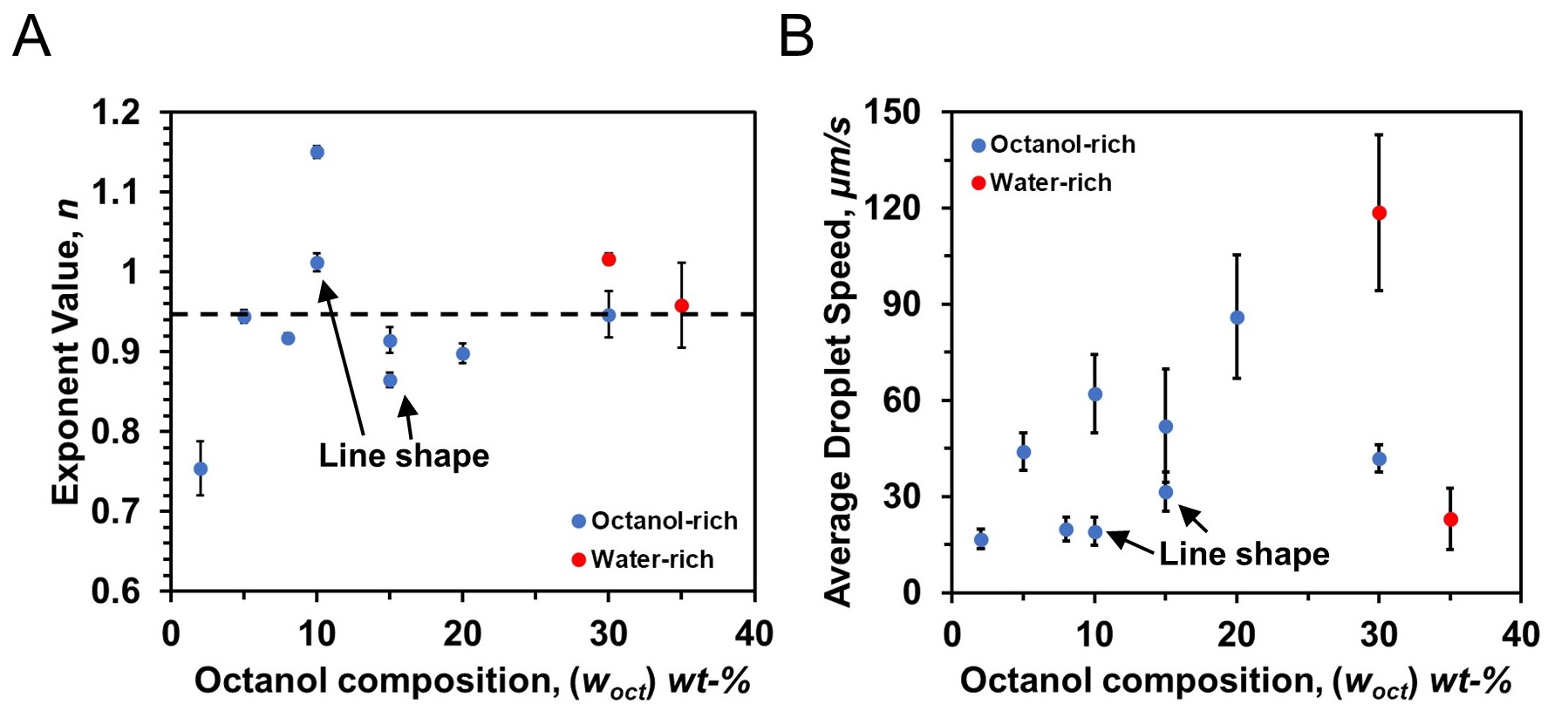}
	\caption{(A) Droplet displacement scaling as a function of octanol wt-\% in solution A. (B) Average droplet speed as a function of octanol wt-\% in solution A. Oil-rich droplets formed from line-shaped protrusions are annotated.
	}
	\label{summary}
\end{figure}

The exponent values obtained from the displacement versus time at octanol compositions in solution A from 2 to 30\% (Compositions 1-9 in Table \ref{Experimental Compositions Table}), presented in Figure \ref{summary}A, were found to scale ($l \sim t^{0.96}$) with an expected uncertainty of $\pm$ $0.10$. However, at 2 wt-\% octanol, the displacement of octanol-rich droplets scaled with ($l \sim t^{0.75}$). At 10 wt-\% octanol composition, the protrusion shape changes from a triangular shape to a line shape where the displacement of droplets scaled with an n value approaching the displacement behaviour of evaporating droplets spreading on a liquid substrate ($l \sim t^{1.2}$) \cite{Wodlei2018}. These scaling results indicate the balance between the Marangoni stress and viscous stress on droplet movement, as was observed for droplet movement enhanced by the Marangoni stress arising from concentration gradients \cite{Rafai2002}.

Figure \ref{summary}B shows the average droplet speed at all positions as a function of octanol composition in solution A from 2 to 35 wt-\%. We have included octanol-rich droplet speeds arising from different boundary shape features, in addition to data on water-rich droplets in the oil-rich zone. The results indicate an overall increasing trend of oil-rich droplet speed from 2 to 20 wt-\% octanol composition, which then decreases at the 30 wt-\% compositions. A maximum value of oil-rich droplet speed is suspected from a composition that intersects the phase boundary close to the plait point, as the trend suggests. At 10 to 20 wt-\% octanol compositions, octanol-rich droplets formed from triangular protrusions and line-shaped protrusions at the boundary location. The results summarized show that droplets from the triangular-shaped protrusions displace faster than those from the line-shaped protrusions. The composition gradient along a line-shaped protrusion is expected to be less steep compared to one from a triangular protrusion. If we consider our previous results (Figure \ref{nozzleconcentrationgradient}G), the larger the angle of the protrusion $\omega$, the larger the propulsion angle $\theta$, and thus the slower the droplet speeds.

\newpage
\section{Conclusions}
We show that oil-rich microdroplets confined in a 2D chamber can travel at speeds up to 150 $\mu m/s$. The motion of oil-rich microdroplets in the water-rich zone is driven by local composition gradients arising from liquid-liquid phase separation during diffusive mixing of water as a poor solvent with a ternary liquid mixture. The fast motion of oil-rich microdroplets induces directional flow in the confined space, mobilizing water-rich microdroplets in the oil-rich zone. For both water-rich microdroplets in an oil-rich zone and oil-rich droplets in a water-rich zone formed during phase separation, the displacement of the microdroplets scales linearly with time. We show the sharpest composition gradient is created at the locations where the droplets form. The speed of droplets is responsive to the shape of the local diffusive mixing boundary. Our findings highlight the possibility of enhanced transport phenomenon via liquid-liquid phase separation in confined spaces. 

\section{Acknowledgement}
This project is supported by the Natural Science and Engineering Research Council of Canada (NSERC) and Future Energy Systems (Canada First Research Excellence Fund). This research was undertaken, in part, thanks to funding from the Canada Research Chairs program.

\printnomenclature

\bibliography{literature}

\end{document}